# Neuromorphic nanocluster networks: Critical role of the substrate in nano-synapse formation

Wenkai Wu[1], Alexey V. Verkhovtsev[2], Theodoros Pavloudis[1,3], Andrey V. Solov'yov[2], Richard E. Palmer[1,*]

[1]Nanomaterials Lab, Faculty of Science and Engineering, Swansea University, Bay Campus, Fabian Way, SA1 8EN, Swansea, UK
[2]MBN Research Center, Altenhöferallee 3, 60438 Frankfurt am Main, Germany
[3]School of Physics, Faculty of Sciences, Aristotle University of Thessaloniki, GR-54124 Thessaloniki, Greece

*Corresponding author. E-mail address: R.E.Palmer@swansea.ac.uk

**Abstract**
Atomic cluster-based networks represent a promising architecture for the realization of neuromorphic computing systems, which may overcome some of the limitations of the current computing paradigm. The formation and breakage of synapses between the clusters are of utmost importance for the functioning of these computing systems. This paper reports the results of molecular dynamics simulations of synapse (bridge) formation at elevated temperature and thermal breaking processes between 2.8 nanometer-sized $Au_{1415}$ clusters deposited on a carbon substrate, a model system. Crucially, we find that the bridge formation process is driven by the diffusion of gold atoms along the substrate, however small the gap between the clusters themselves. The complementary simulations of the bridge breaking process reveal the existence of a threshold bias voltage to activate bridge rupture via Joule heating. These results provide an atomistic-level understanding of the fundamental dynamical processes occurring in neuromorphic cluster arrays.

1. Introduction

Existing computing technologies based on the traditional von Neumann paradigm may now be approaching their fundamental limits and require an increasing amount of energy resources [1–3]. Partly driven by this problem, a new interdisciplinary research area called neuromorphic computing has appeared recently [4], being motivated by the incredible capability of the human brain to perform specific tasks, like pattern recognition, in a very energy-efficient way [5–7]. Neuromorphic computing aims to mimic the function of a human brain by assembling artificial neurons and synapses [8] and is considered promising to overcome some of the limitations of conventional computers [9–11].

A proposed architecture for neuromorphic computing is the assembly of interconnected, synapse-like switching devices [11–13] such as memristors, which change their resistance depending on the history of the bias voltage applied to the system [14,15]. Switching behaviour in such devices can be achieved due to the dynamic formation and breakage of nanoscale conductive filaments [16]. Metal cluster films fabricated by the gas-phase deposition technique have been reported to show similar non-ohmic switching behaviour [17–22], therefore networks of atomic clusters and small nanoparticles can function as neuromorphic network building blocks [23]. Benefiting from the developing cluster deposition technologies [24–26], this approach is simpler and possibly cheaper than other techniques, such as lithographic methods [27–29], as employed to create robust intra- and inter-device connections in a deterministic way.

The mechanisms of formation and breakage of the conductive connections between clusters have been discussed in the literature. In connection with experimental observations, notably conductance measurements [18,30], the electric field-induced surface diffusion and evaporation and the van der Waals interaction between metal atoms are considered the main drivers of the formation of atomic-scale connections between the clusters [31–33]. After the formation of the connection, if no bias is applied, the width of the connection increases with time following a power-law function, as shown in experimental observation by scanning transmission electron microscopy (STEM) and simulation results using the kinetic Monte Carlo approach [34,35]. The exponent's value can be attributed to the combined effects of diffusion caused by surface tension and the viscosity of the material [36,37]. On the other hand, the breakage of the connections is explained by electromigration introduced by the electric current running through the connections. Numerous studies have emphasized the contribution of electromigration to the breakage or resistive switching of metal nanowires (NWs) [38–41].

An alternative mechanism for the breakage is the Joule heating induced by the current running through the connections [17]. High temperatures up to thousands of kelvin could be induced by small voltage drops on the order of ~100 mV when the size of a point contact in a complex nanoscale junction is smaller than the mean free path of electrons [42]. The thermally induced breaking of NWs has been frequently reported [43–45], and relevant computer simulations have been carried out [46,47]. Our prior research has demonstrated that Joule heating can break the nanofilaments connecting the clusters [48]. Strong local heating can be observed in a NW network when current flows [49]. The breaking process of the NW by heat generation can be observed in situ using transmission electron microscopy [50]. Meanwhile, multiple quantum conductance models and related experiments and simulations have been employed to evaluate the current through a filament [51–55]. These studies have established a foundation for evaluating the Joule heat generation in a nanofilament based on the bias voltage applied.

Although various experimental studies and computer simulations treat neuromorphic cluster dynamics on various substrates [56–58], previous studies on the mechanism of connection formation and breakage have typically focused merely on the clusters themselves [31–33]. The role of the substrate in the formation and breakage of conductive filaments remains unresolved and needs further exploration.

In this paper, the formation and breaking of a nanoscale conductive filament between nanometer-sized gold clusters placed on a carbon substrate are studied by means of molecular dynamics (MD) simulations. Our result shows that the graphite substrate serves as a pathway for the clusters to diffuse along, and the formation and breakage of the connection always happen along the substrate. The characteristic times for bridge formation by diffusion of atoms along the substrate are evaluated at elevated temperatures (T = 500-800 K) for different cluster orientations and distances between the clusters and used to estimate the bridge formation time under experimentally relevant conditions. After the formation of the bridge, the heat generated by the electric current for different bias voltage values is estimated using a model approach for determining the conductance of nanoscopic systems, and the nanofilament breaking process on the substrate under this thermal stimulus has been simulated. The presented results provide an atomistic-level understanding of the fundamental processes involving the conductive connections between deposited metal clusters. Our results highlight the important role of the substrate in the bridge formation and breakage processes, which offers new insight into the fundamental processes in cluster-based neuromorphic computing systems.

## 2. Computational methodology

The MD simulations were performed with the MBN Explorer software package [59]. The creation of the systems and analysis of the results was performed with its accompanying multi-task toolkit, MBN Studio [60].

The studied system is a cuboctahedral Au$_{1415}$ cluster placed on top of a four-layer thick graphite substrate, see Figure 1. The cuboctahedral structural motif was chosen based on the presence of both (100) and (111) facets of comparable total area. Au$_{1415}$ cuboctahedral nanoparticles (NPs) comprise six rectangular 8x8-atom (100) facets and eight equilateral triangular 6-atom per side (111) facets. This facilitates the comparison between different orientations using the same model as a reference. Carbon substrates (either graphitic or amorphous carbon) are easily available and have been widely used for atomic cluster deposition. Gold has also been a model material in nanocluster science for decades. Therefore, the cluster-support combination considered in this paper is a system that can be easily replicated in future experimental observations.

The system was simulated using periodic boundary conditions so that connecting bridges were formed between the cluster and its neighbouring periodic images. Four different Au$_{1415}$ geometry arrangements were simulated to demonstrate the impact of different cluster orientations: (i) the (100) facet of the gold cluster was on the graphite layer (labelled as Au$_{(100)}$); (ii) rotation of the cluster by 45° along the z-axis based on (i) (Au$_{(100)rot}$); (iii) the (111) facet of the cluster was on the graphite layer (Au$_{(111)}$); and (iv) rotation of the cluster by 60° along the z-axis based on (iii) (Au$_{(111)rot}$). These geometrical arrangements are shown in Figure 1 (b-e) respectively. The distance between the cluster and its periodic image was varied by adjusting the simulation box size. The atoms of bottommost layer of the substrate were fixed in place to avoid translational motion of the whole system.

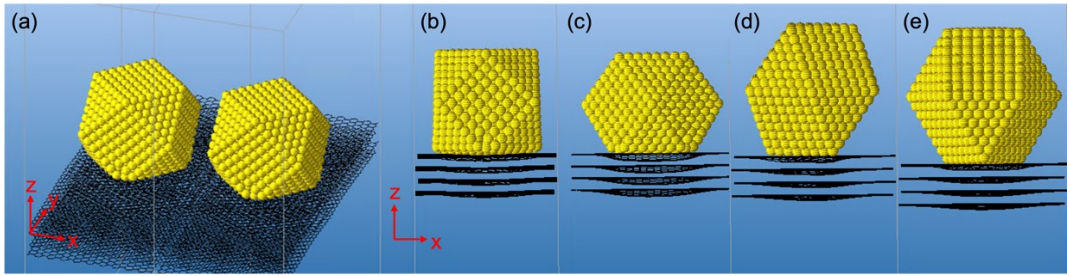

**Figure 1.** The geometry of the simulated system. A cuboctahedral Au$_{1415}$ cluster is placed on top of a four-layer thick graphite substrate and periodic boundary conditions are employed. **(a)** Snapshot of the system and its periodic image. **(b)** The (100) facet of the gold cluster is on the graphite layer (Au$_{(100)}$). **(c)** Based on Au$_{(100)}$, the cluster is rotated 45° around the z-axis (Au$_{(100)rot}$). **(d)** The (111) facet of the cluster is on the graphite layer (Au$_{(111)}$). **(e)** Based on Au$_{(111)}$, the cluster is rotated 60° along around the z-axis (Au$_{(111)rot}$).

The many-body Gupta potential [61,62] was used to model the interaction between gold atoms. The bond-order Brenner potential [63] was used to describe the interaction between covalently bonded carbon atoms within each graphite layer, while a Lennard-Jones potential was employed to account for the van der Waals interaction between the graphite layers and also the Au-C interactions. The parameters for these interatomic potentials are listed in Ref. [57].

In the simulation of bridge formation, the system's temperature was controlled by means of a Langevin thermostat with a damping time of 1 ps; the integration time step was set equal to 2 fs. After the initial energy minimization and relaxation of each system at 300 K, a series of constant-temperature MD simulations were carried out for each geometry and distance between the cluster and its adjacent images within the temperature range from 500 to 800 K. The total duration of each simulation was 10 ns. Importantly, we have not observed any rearrangements in the structure of cuboctahedral clusters during the initial relaxation and equilibration stages.

The initial geometries for the simulations of bridge breaking were taken from the outputs of the bridge formation simulations. Prior to the simulations of bridge breaking, the systems were equilibrated at 300 K for 100 ps using a Langevin thermostat. Gold atoms that appeared in the gap between the neighboring clusters in the initial Au$_{(100)}$ geometry (see Figure 1 (a)) were assigned to the bridge region. The model proposed by Wexler [52,55] was used to estimate the conductance of the bridge and the current running through the system for a specific bias voltage. The heating rate for the system due to the Joule heating mechanism was then evaluated according to the estimated current and conductance. The velocities of gold atoms in the bridge region were rescaled according to the evaluated heating rate. The system then evolved over 2 ps without a thermostat, and the last frame of the simulation was used to re-evaluate the conductance and current for the chosen bias voltage. Multiple subsequent simulations were performed in this way for a total simulation time of 100 ps. The simulations were carried out for the values of bias voltage in the range V = 10 ~ 100 mV applied between the cluster and its neighboring image, which corresponds to the typical cluster sizes and distances between the electrodes in experiments [18,34]. Although the realistic arrangement of clusters in a randomly assembled cluster film is, of course, more complex than the presented model, the chosen cluster geometries, as well as the temperature and bias voltage ranges, are representative of experiments in this research field. This permits us to explore at the atomistic scale the mechanisms which drive the function of such a device.

### 3. Results and discussion

*3.1. Bridge formation process*

The process of bridge formation between two gold $Au_{1415}$ clusters deposited on a carbon substrate is illustrated in Figure 2. The process starts with the diffusion of gold atoms from the clusters along the substrate, leading to a narrowing of the gap between the clusters. Then, a thin bridge with the width of one or two atoms is formed so that the clusters become connected. Figure 2(a) shows a schematic illustration of this process, and Figure 2(d) shows the corresponding snapshot of an MD simulation performed at 700 K. The width of the bridge then increases to about half of the cluster's diameter (equal to ~2.8 nm) within the first ~200 ps of the simulation. The height of such a bridge corresponds to one monolayer of gold atoms. The corresponding schematic image and simulation snapshot are shown in Fig. 2(b) and Fig. 2(e), respectively. The increase of the width then becomes slower, and a second layer of atoms appears on the bridge. The system's geometries obtained at the end of 1 ns-long simulations performed at 700 K are shown in Fig. 2(c) and Fig. 2(f). It is notable that in the simulations carried out in this study, the bridge was always formed along the substrate for any initial distance between the clusters above 5 Å and for any cluster orientations. Figure 2(g) shows the initial $Au_{(100)rot}$ cluster orientation, where the shortest distance between atoms of two neighboring clusters is equal to 7 Å, a value smaller than the chosen cutoff distance for Au-Au interactions. Despite this closeness, the atoms still diffuse along the substrate and form a bridge on the substrate, as shown in Figure 2(h).

Due to the limited simulation time (10 ns), the bridge formation was not observed in the simulations performed at room temperature. Therefore, higher temperatures in the range 500-800 K have been considered, and a relation between the bridge formation time (defined as the time needed for the formation of the monolayer bridge shown in Fig. 2(a)) and the thermostat temperature was established. Figure 3 shows the dependence of bridge formation time on the distance between the clusters. The figure shows the simulation results for temperatures above 550 K at which the bridge was formed in every simulation run. Dots indicate the results of each run, and crosses illustrate the average bridge formation time for a specific temperature. The formation time is shown on a logarithmic scale.

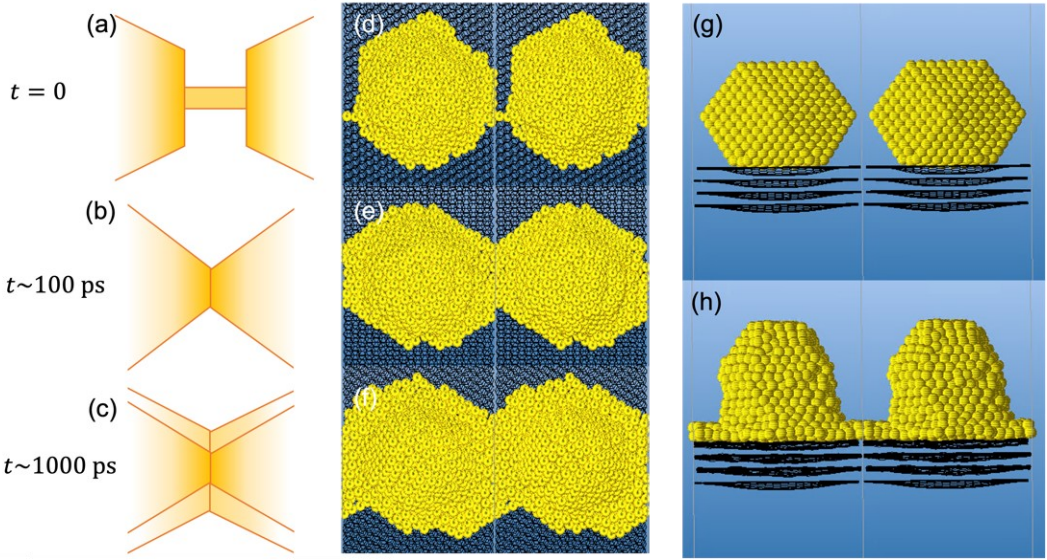

**Figure 2.** A typical bridge formation process. (a-c) Sketch of the bridge formation process as viewed from above. **(a)** A bridge one or two atoms in width is formed between two neighboring Au$_{1415}$ clusters. **(b)** The bridge increases in width over the ~100 ps of the simulation time. **(c)** The lateral expansion of the bridge slows, and a second layer of atoms is formed on top. **(d-f)** Corresponding top views of the simulated MD trajectories at a temperature of 700 K. **(g-h)** Side view of the simulation outcome for the Au$_{(100)rot}$ geometry arrangement: initial configuration **(g)** and steady state after the bridge has formed **(h)**. The bridge formation always takes place along the substrate and not between the initially closest point between two clusters.

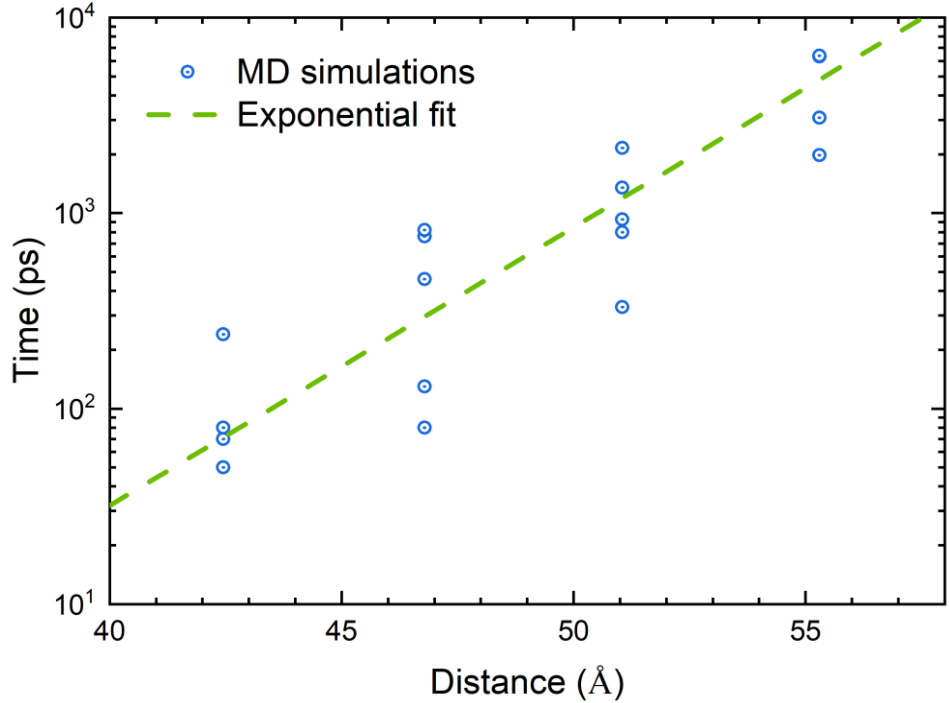

**Figure 3.** Bridge formation time as a function of the distance between the centers of neighboring clusters at T = 750 K.

The formation of nanofilaments has been shown to be highly affected by surface atomic self-diffusion [64]. According to surface diffusion theory, the rate at which a single particle (e.g. an atom) diffuses on a surface is given by:

$$\Gamma = \frac{1}{\Delta t} = \nu \exp\left(-\frac{E_a}{k_B T}\right) \qquad (1)$$

where $E_a$ is the diffusion activation energy defined as the value of the potential barrier that should be overcome by an atom to occupy a neighboring position on the surface, $v$ is the attempt rate, $T$ is the temperature of the system, $k_B$ is the Boltzmann constant, and $\Delta t$ is the characteristic time for particle diffusion on the said surface[65–67]. The calculation of the diffusion activation energy is a separate research task that goes beyond the scope of this study. However, we note that the diffusion activation energy for gold atoms diffusing on graphite was calculated in several earlier studies by means of density functional theory and molecular dynamics simulations. The values of $E_a$ reported in these studies are in the range 0.05 – 0.10 eV[68,69]. An important outcome from Eq. (1) is that $\Delta t$ depends exponentially on $1/T$.

If one thinks of the bridge formation process as an accumulation of the diffusive steps of all the atoms, it is reasonable to consider the bridge formation time to be governed by an exponential relation with $1/T$. Indeed, our simulations performed at different temperatures result in similar systems' geometries, with the only variable being the bridge formation time. A fit of our data with the exponential function given by Eq. 1 predicts a bridge formation time of ~10 µs at T = 300 K (see Figure 4). Therefore, we believe that surface diffusion can be the dominant mechanism of the bridge formation process for the systems and the temperature range considered in this study.

An alternative to the single adatom diffusion process for the explanation of bridge formation can be the coordinated diffusion of a group of atoms. In Refs. [70] and [71], the growth and coalescence mechanisms of Au NPs of sizes similar to the ones used in this work were found to be the result of "concerted displacements of many atoms".

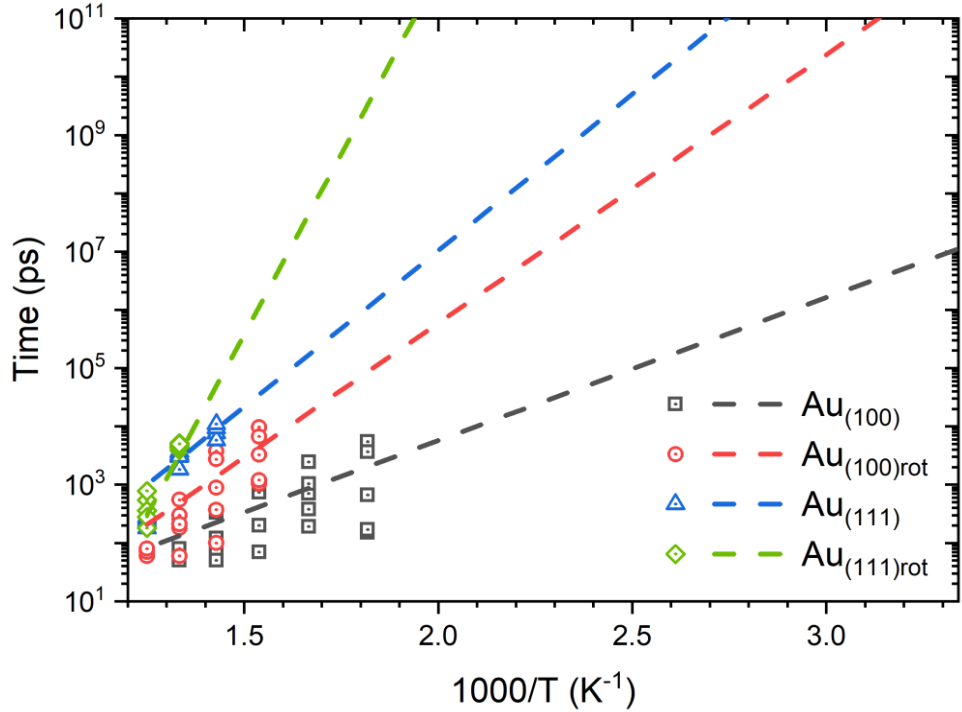

**Figure 4.** The dependence of bridge formation time on temperature for different geometric orientations shown in Figure 1. Symbols show the results of independent MD simulations; dashed lines illustrate the exponential fit.

Figure 4 shows the dependence of bridge formation time on temperature for different cluster orientations. The results of MD simulations (symbols) were fitted for each cluster orientation with an exponential function (lines). The distance between neighbouring clusters was set to 7 Å for all configurations. Interestingly, we did not observe a bridge formation of $Au_{(111)}$ and $Au_{(111)rot}$ orientation (blue triangles and green diamonds, respectively) at temperatures below 750 K over a simulation time of 10 nanoseconds. The predicted formation time at 300 K is ~10 µs for $Au_{(100)}$ orientation, ~1 s for Au$_{(100)rot}$, and over one year for Au$_{(111)}$ and Au$_{(111)rot}$ orientations. A reason for the significantly faster formation process for the Au$_{(100)}$ orientation may be that the (100) facets contain more atoms compared with the (111) facets of these clusters (64 and 36 atoms, respectively, see Figure 1). Additionally, the (111) facet of gold is a good match to the hexagonal graphitic structure[72], a factor tending to the stabilization of the cluster. In general, the predicted formation times for all orientations at 300 K show a large variation, from ~µs to years, which reflects the experimental complexity of the switching behaviour in randomly assembled cluster films [18,22].

After the bridge formation, the neck width (defined as the width at the middle of the bridge) continues growing as shown in Figure 2. The "neck width – time" relation for Au$_{(100)}$ cluster orientation at different temperatures is shown in Figure 5(a). Eight to ten independent simulations have been carried out at each temperature, and each symbol shows the average time when the bridge reaches a certain width. The data agree well with a power law $D \sim t^\alpha$, where $D$ is the neck width of the bridge and $t$ is time [34,35]. The relation between the power $\alpha$ and the temperature is shown in the inset. It follows from the performed simulations that $\alpha = 0.34 \pm 0.03$ and it is practically independent of the temperature.

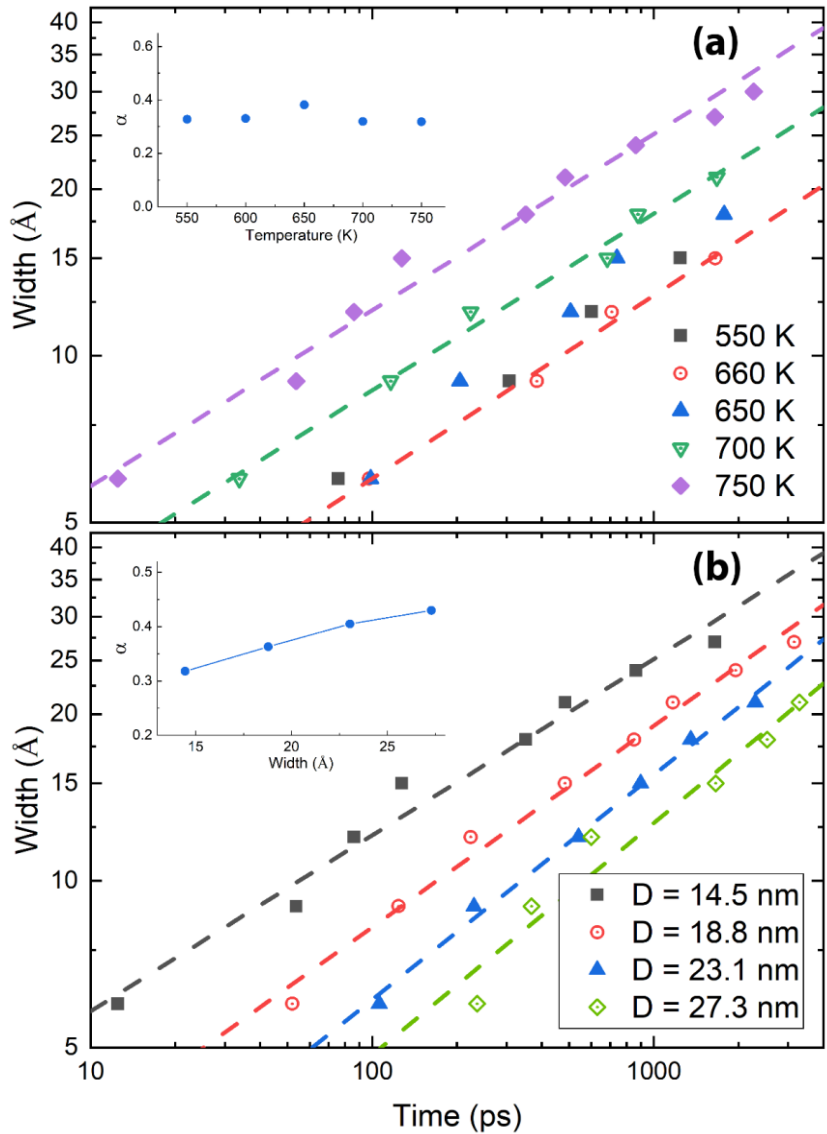

**Figure 5.** Width - time relation of the bridge for **(a)** different temperatures and **(b)** different distances between clusters. Each symbol shows the average time when the bridge reaches certain width. Dashed lines represent fits using the power law function $D \sim t^\alpha$. The insets show the relations between the power $\alpha$ and the temperature (panel (a)) or the distance between clusters (panel (b)).

A similar analysis was done as a function of the distance between the neighbouring clusters. Figure 5(b) shows the neck width – time relation for different neighbouring distances at 750 K. The power law $D \sim t^\alpha$ is also used for fitting the data and the relation between the power $\alpha$ and the distance is shown in the inset. In this case, a linear increase of $\alpha$ with the distance can be observed. The growth of the neck width has been experimentally observed previously [35]. Monte Carlo simulations have attempted to shed light onto the underlying mechanism [34,35]. It was shown that the neck radius $r$ between two coalescing microscale solid particles follows a power law $r \propto t^\alpha$, where $\alpha$ is specific to the physical coalescence process and lies between 1/6 (surface diffusion) [36] and 1/2 (viscous flow) [37]. Our results shown in Figure 5(a) agree well with the power law behaviour, as discussed, and the power $\alpha = 0.34 \pm 0.03$ is in the range of these previous simulation results and very close to the in-situ observation result in Ref. [35] where the power $\alpha \sim 0.32$. Our finding that the exponent $\alpha$ is independent of temperature is also compatible with the previous simulation predictions. Note that in Refs. [34,73] the clusters were spherical, and no substrate was considered, so the cross-section of the neck was circular. In the present simulations, the neck has a bow-tie shape with a thickness of one or two atomic layers (see Figure 2(e,f,h)). However, it can be seen that despite the geometry difference caused by the surface interaction between the gold and the substrate, the neck growth mechanism remains unchanged. On the other hand, our result in Figure 6 shows that the mechanism is affected by the initial distance between the clusters, but the value of $\alpha$ is still within the range from 1/6 to 1/2. A longer initial distance makes the diffused cluster atoms deviate more from the circular shape when the bridge is formed. $\alpha$ is approaching 1/2 for a longer initial distance, which shows that the viscosity flow mechanism becomes dominant when the geometry of the system deviates from the ideal circular shape.

*3.2. Bridge breaking process*

The bridge breaking process under different values of bias voltage, from 10 to 100 mV, at T = 300 K was also studied. The geometry of the bridge with a width of one or two atoms was chosen to be the initial state because the current starts to run through the bridge as soon as the bridge is formed, as described in previous experimental works [17–22]. The conductance of the bridge is evaluated using the theoretical model suggested by Wexler [52]:

$$G = \frac{k_F^2}{4\pi} S G_0 \left[1 + \frac{3\pi}{8} \Gamma(K) \frac{r}{l}\right]^{-1} \qquad (2)$$

where $k_F$ is the Fermi wavevector (in our case, gold, $k_F = 1.204$ Å$^{-1}$, see Ref. [74]), $G_0 = 2e^2/h$ is the quantum conductance unit, $r$ is the radius of the cross section of the bridge (in our case is the van der Waals radius of gold atom 1.44 Å), $l$ is the electron mean free path, $S$ is the cross-sectional area of the bridge (in our case $S = \pi r^2$), $\Gamma$ is a slowly varying function of order ~1 which can be seen as constant in our case, and $K = l/r$ is the so-called Knudsen number (the ratio of mean free path and the representative physical length scale, in our case, the radius of the cross section of the bridge). Experimental results show a "conductance – surface area" relation between the gold link formed by mechanical stretch [55]. The data was fitted well with Wexler's expression. The fitted values $\Gamma(K) = 0.7$ and $l = 38$ Å. The conductance of the bridge was calculated with the information above, and then the current and power were evaluated for different values of bias voltage.

At the bias voltage values above 46 mV, gold atoms in the bridge region will have intense thermally-induced atomic motion and eventually break the bridge at t~25 ps. Local heat is generated due to different heat transfer properties between the cluster and the substrate. The temperature of the cluster reaches ~700 K when the breakage happens. With a bias lower than 30 mV, the atomic movement in the bridge part is not intense enough to break the bridge. The temperature of the cluster keeps increasing, and the width of the bridge increases, similar to the bridge formation process. Figure 6 illustrates the different results in the case of high and low bias. The estimated conductance, current and power of heat generation are shown in Table 1. The conductance $G$ is given in the units of quantum conductance $G_0 = 2e^2/h$.

Table 1: Characteristic parameters of the bridge

| Conductance | Bias (mV) | Current (µA) | Power (nW) |
| --- | --- | --- | --- |
| $0.72 G_0$ | 46 | 2.6 | 120 |
|  | 30 | 1.7 | 50 |

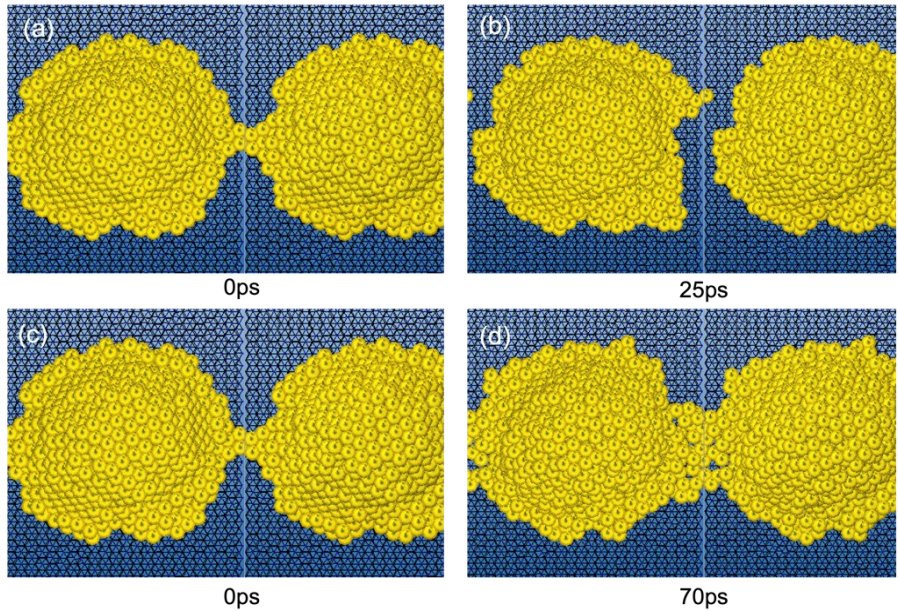

**Figure 6.** Results of bridge breaking simulation for different applied bias voltages. (a), (b) are the initial and breaking frames of a simulation of 46 mV. (c), (d) are the initial frame and intermediate frame of a simulation of 30 mV. The time of the corresponding frame is indicated below.

Our results indicate that, at the applied bias voltage higher than 46 mV, the bridge between the neighbouring clusters breaks during the first 20-30 ps and then reforms again on the time scale of about 100 ps. After the configuration shown in Fig. 6(b) is formed, there is a dynamic interplay between the bridge breaking and formation processes over the time scale of about 10 ps. After that, a stable bridge is formed again, becoming thicker as the simulation time increases. In contrast, the bridge does not break and continues to grow when the bias is lower than 30 mV. A bias between these two values will result in random growth or breaking behaviour. Therefore, our results indicate that a certain bias level is required to break the bridge, but the threshold is obscure. Several experimental reports mentioned the existence of a threshold [17,18] and the complexity of the switching behaviour near the threshold has been reported via conductance measurements [18–20]. Our simulation results reflect these features and can act as a reference for future studies.

4. Conclusions

In summary, aiming to provide an understanding of the formation and breakage of the conductive filamentary bridges between neighbouring clusters in neuromorphic devices, we built a series of model systems and performed MD simulations on the formation and breaking process of the bridges between clusters on a graphite substrate.

In the bridge forming simulations, a high temperature was used to accelerate the process and provide results within the MD simulation time, and the time scale at room temperature was extrapolated from fittings of a series of these higher temperature results. The diffusion of atoms on graphite substrate contributes greatly to the formation mechanism in our simulations. In the bridge breaking simulations, the conductance of the bridge was estimated by the quantum conductance model, and Joule heating under particular bias conditions was calculated. The temperature variation was applied by re-scaling the speed of atoms on the bridge, and the breaking process was revealed by the simulation. Our results indicate the validity of the Joule heating mechanism in the breaking process of the bridge. We also show the existence of a bias threshold to activate switching behaviour and demonstrate the complexity of the switching process in experiments.

These atomistic simulations provide valuable insight into the mechanisms that govern the behaviour of cluster-based neuromorphic systems. Our models took into account the clusters and the support. Parameters of the model have been chosen according to the experimental conditions, making our results representative and a good reference for further studies. New mechanisms, such as diffusion and different formation and breakage process, were revealed by our model. However, since more complicated and asymmetric situations can exist in a real cluster network, more research is needed to present a full picture of the mechanism in the forming and breaking behavior. We hope this work will contribute to a better understanding of nanocluster networks.

Follow-up works must include multiple nanoclusters, of varying orientations, deposited on a series of substrates. Au binds weakly on graphene, with a small binding energy and large equilibrium separations [75]. A future work might investigate substrates that show strong binding with Au. The comparison of the results with the present work would be of great interest for the realization of cluster-based neuromorphic devices.

**Acknowledgements**
The authors are grateful for partial financial support by the European Commission through the RADON project (GA 872494) within the H2020-MSCA-RISE-2019 call. This article is also based upon work from the COST Action CA20129 MultIChem, supported by COST (European Cooperation in Science and Technology). TP acknowledges Swansea University support via project 74143 of the Research Committee of AUTh. This work was also supported in part by Deutsche Forschungsgemeinschaft (Project no. 415716638). The authors acknowledge the computational support by the Goethe-HLR cluster of the Frankfurt Center for Scientific Computing and by the Supercomputing Wales project, which is part-funded by the European Regional Development Fund (ERDF) via the Welsh Government.

**References**
[1] Waldrop, M. M. The Chips Are down for Moore's Law. *Nature* **2016**, *530* (7589), 144–147.
[2] UN. A New Circular Vision for Electronics Time for a Global Reboot. *World Econ. Forum* **2019**, No. URL: https://www.weforum.org/reports/a-new-circula.
[3] Nawrocki, R. A.; Voyles, R. M.; Shaheen, S. E. A Mini Review of Neuromorphic Architectures and Implementations. *IEEE Trans. Electron Devices* **2016**, *63* (10), 3819–3829.
[4] Jaeger, H. Towards a Generalized Theory Comprising Digital, Neuromorphic and Unconventional Computing. *Neuromorphic Comput. Eng.* **2021**, *1* (1), 012002.
[5] Bullmore, E.; Sporns, O. The Economy of Brain Network Organization. *Nat. Rev. Neurosci.* **2012**, *13* (5), 336–349.
[6] Wang, Z.; Wu, H.; Burr, G. W.; Hwang, C. S.; Wang, K. L.; Xia, Q.; Yang, J. J. Resistive Switching Materials for Information Processing. *Nat. Rev. Mater.* **2020**, *5* (3), 173–195.
[7] Torrejon, J.; Riou, M.; Araujo, F. A.; Tsunegi, S.; Khalsa, G.; Querlioz, D.; Bortolotti, P.; Cros, V.; Yakushiji, K.; Fukushima, A.; Kubota, H.; Yuasa, S.; Stiles, M. D.; Grollier, J. Neuromorphic Computing with Nanoscale Spintronic Oscillators. *Nature* **2017**, *547* (7664), 428–431.

[8] Mead, C. Neuromorphic Electronic Systems. *Proc. IEEE* **1990**, *78* (10), 1629–1636.
[9] Anderson, H. C. Neural Network Machines. *IEEE Potentials* **1989**, *8* (1), 13–16.
[10] Rozenberg, G.; Back, T.; Kok, J. N. *Handbook of Natural Computing*; Springer Berlin Heidelberg, 2012; Vol. 1–4.
[11] Avizienis, A. V.; Sillin, H. O.; Martin-Olmos, C.; Shieh, H. H.; Aono, M.; Stieg, A. Z.; Gimzewski, J. K. Neuromorphic Atomic Switch Networks. *PLoS One* **2012**, *7* (8), e42772.
[12] Ohno, T.; Hasegawa, T.; Tsuruoka, T.; Terabe, K.; Gimzewski, J. K.; Aono, M. Short-Term Plasticity and Long-Term Potentiation Mimicked in Single Inorganic Synapses. *Nat. Mater.* **2011**, *10* (8), 591–595.
[13] Stieg, A. Z.; Avizienis, A. V.; Sillin, H. O.; Martin-Olmos, C.; Aono, M.; Gimzewski, J. K. Emergent Criticality in Complex Turing B-Type Atomic Switch Networks. *Adv. Mater.* **2012**, *24* (2), 286–293.
[14] Wuttig, M.; Yamada, N. Phase-Change Materials for Rewriteable Data Storage. *Nat. Mater.* **2007**, *6* (11), 824–832.
[15] Chua, L. O. Memristor—The Missing Circuit Element. *IEEE Trans. Circuit Theory* **1971**, *18* (5), 507–519.
[16] Prodromakis, T.; Toumazou, C.; Chua, L. Two Centuries of Memristors. *Nat. Mater.* **2012**, *11* (6), 478–481.
[17] Mirigliano, M.; Decastri, D.; Pullia, A.; Dellasega, D.; Casu, A.; Falqui, A.; Milani, P. Complex Electrical Spiking Activity in Resistive Switching Nanostructured Au Two-Terminal Devices. *Nanotechnology* **2020**, *31* (23), 234001.
[18] Sattar, A.; Fostner, S.; Brown, S. A. Quantized Conductance and Switching in Percolating Nanoparticle Films. *Phys. Rev. Lett.* **2013**, *111* (13), 136808.
[19] Mirigliano, M.; Borghi, F.; Podestà, A.; Antidormi, A.; Colombo, L.; Milani, P. Non-Ohmic Behavior and Resistive Switching of Au Cluster-Assembled Films beyond the Percolation Threshold. *Nanoscale Adv.* **2019**, *1* (8), 3119–3130.
[20] Bose, S. K.; Mallinson, J. B.; Gazoni, R. M.; Brown, S. A. Stable Self-Assembled Atomic-Switch Networks for Neuromorphic Applications. *IEEE Trans. Electron Devices* **2017**, *64* (12), 5194–5201.
[21] Mallinson, J. B.; Shirai, S.; Acharya, S. K.; Bose, S. K.; Galli, E.; Brown, S. A. Avalanches and Criticality in Self-Organized Nanoscale Networks. *Sci. Adv.* **2019**, *5* (11), eaaw8483.
[22] Minnai, C.; Mirigliano, M.; Brown, S. A.; Milani, P. The Nanocoherer: An Electrically and Mechanically Resettable Resistive Switching Device Based on Gold Clusters Assembled on Paper. *Nano Futur.* **2018**, *2* (1), 1002.
[23] Minnai, C.; Bellacicca, A.; Brown, S. A.; Milani, P. Facile Fabrication of Complex Networks of Memristive Devices. *Sci. Rep.* **2017**, *7* (1), 7955.
[24] Wegner, K.; Piseri, P.; Tafreshi, H. V.; Milani, P. Cluster Beam Deposition: A Tool for Nanoscale Science and Technology. *J. Phys. D. Appl. Phys.* **2006**, *39* (22), 439–460.
[25] Zhang, J. J.; Sun, H. J.; Li, Y.; Wang, Q.; Xu, X. H.; Miao, X. S. AgInSbTe Memristor with Gradual Resistance Tuning. *Appl. Phys. Lett.* **2013**, *102* (18), 183513.
[26] Palmer, R. E.; Cao, L.; Yin, F. Note: Proof of Principle of a New Type of Cluster Beam Source with Potential for Scale-Up. *Rev. Sci. Instrum.* **2016**, *87* (4), 10–13.
[27] Joshua Yang, J.; Miao, F.; Pickett, M. D.; Ohlberg, D. A. A.; Stewart, D. R.; Lau, C. N.; Williams, R. S. The Mechanism of Electroforming of Metal Oxide Memristive Switches. *Nanotechnology* **2009**, *20* (21), 215201.
[28] Du, C.; Cai, F.; Zidan, M. A.; Ma, W.; Lee, S. H.; Lu, W. D. Reservoir Computing Using Dynamic Memristors for Temporal Information Processing. *Nat. Commun.* **2017**, *8*, 2204.
[29] Bose, S. K.; Shirai, S.; Mallinson, J. B.; Brown, S. A. Synaptic Dynamics in Complex Self-Assembled Nanoparticle Networks. *Faraday Discuss.* **2019**, *213*, 471–485.
[30] Pike, M. D.; Bose, S. K.; Mallinson, J. B.; Acharya, S. K.; Shirai, S.; Galli, E.; Weddell, S. J.; Bones, P. J.; Arnold, M. D.; Brown, S. A. Atomic Scale Dynamics Drive Brain-like Avalanches in Percolating Nanostructured Networks. *Nano Lett.* **2020**, *20* (5), 3935–3942.
[31] Olsen, M.; Hummelgård, M.; Olin, H. Surface Modifications by Field Induced Diffusion. *PLoS One* **2012**, *7* (1), 30106.
[32] Tsong, T. T. Effects of an Electric Field in Atomic Manipulations. *Phys. Rev. B* **1991**, *44* (24),

[33] Mayer, T. M.; Houston, J. E.; Franklin, G. E.; Erchak, A. A.; Michalske, T. A. Electric Field Induced Surface Modification of Au. *J. Appl. Phys.* **1999**, *85* (12), 8170–8177.

[34] Convers, P. Y.; McCarthy, D. N.; Sattar, A.; Natali, F.; Hendy, S. C.; Brown, S. A. Electrical Signature of Nanoscale Coalescence in a Percolating Bi Nanocluster Film. *Phys. Rev. B - Condens. Matter Mater. Phys.* **2010**, *82* (11), 1–5.

[35] Lim, T. H.; McCarthy, D.; Hendy, S. C.; Stevens, K. J.; Brown, S. A.; Tilley, R. D. Real-Time TEM and Kinetic Monte Carlo Studies of the Coalescence of Decahedral Gold Nanoparticles. *ACS Nano* **2009**, *3* (11), 3809–3813.

[36] Kuczynski, G. C. Study of the Sintering of Glass. *J. Appl. Phys.* **1949**, *20* (12), 1160–1163.

[37] Nichols, F. A.; Mullins, W. W. Morphological Changes of a Surface of Revolution Due to Capillarity-Induced Surface Diffusion. *J. Appl. Phys.* **1965**, *36* (6), 1826–1835.

[38] Kim, T. H.; Zhang, X. G.; Nicholson, D. M.; Evans, B. M.; Kulkarni, N. S.; Radhakrishnan, B.; Kenik, E. A.; Li, A. P. Large Discrete Resistance Jump at Grain Boundary in Copper Nanowire. *Nano Lett.* **2010**, *10* (8), 3096–3100.

[39] Johnson, S. L.; Sundararajan, A.; Hunley, D. P.; Strachan, D. R. Memristive Switching of Single-Component Metallic Nanowires. *Nanotechnology* **2010**, *21* (12), 125204.

[40] Song, T. Bin; Chen, Y.; Chung, C. H.; Yang, Y.; Bob, B.; Duan, H. S.; Li, G.; Tu, K. N.; Huang, Y. Nanoscale Joule Heating and Electromigration Enhanced Ripening of Silver Nanowire Contacts. *ACS Nano* **2014**, *8* (3), 2804–2811.

[41] Hoffmann-Vogel, R. Electromigration and the Structure of Metallic Nanocontacts. *Appl. Phys. Rev.* **2017**, *4* (3), 031302.

[42] Halbritter, A.; Csonka, S.; Kolesnychenko, O. Y.; Mihály, G.; Shklyarevskii, O. I.; Shklyarevskii, O. I.; van Kempen, H. Connective Neck Evolution and Conductance Steps in Hot Point Contacts. *Phys. Rev. B - Condens. Matter Mater. Phys.* **2002**, *65* (4), 454131–454138.

[43] Wen, Y. H.; Zhang, Y.; Zheng, J. C.; Zhu, Z. Z.; Sun, S. G. Orientation-Dependent Structural Transition and Melting of Au Nanowires. *J. Phys. Chem. C* **2009**, *113* (48), 20611–20617.

[44] Volk, A.; Knez, D.; Thaler, P.; Hauser, A. W.; Grogger, W.; Hofer, F.; Ernst, W. E. Thermal Instabilities and Rayleigh Breakup of Ultrathin Silver Nanowires Grown in Helium Nanodroplets. *Phys. Chem. Chem. Phys.* **2015**, *17* (38), 24570–24575.

[45] Schnedlitz, M.; Lasserus, M.; Knez, D.; Hauser, A. W.; Hofer, F.; Ernst, W. E. Thermally Induced Breakup of Metallic Nanowires: Experiment and Theory. *Phys. Chem. Chem. Phys.* **2017**, *19* (14), 9402–9408.

[46] Wen, Y. H.; Zhu, Z. Z.; Zhu, R.; Shao, G. F. Size Effects on the Melting of Nickel Nanowires: A Molecular Dynamics Study. *Phys. E Low-Dimensional Syst. Nanostructures* **2004**, *25* (1), 47–54.

[47] Moskovkin, P.; Panshenskov, M.; Lucas, S.; Solov'yov, A. V. Simulation of Nanowire Fragmentation by Means of Kinetic Monte Carlo Approach: 2D Case. *Phys. Status Solidi Basic Res.* **2014**, *251* (7), 1456–1462.

[48] Wu, W.; Pavloudis, T.; Verkhovtsev, A. V; Solov'yov, A. V; Palmer, R. E. Molecular Dynamics Simulation of Nanofilament Breakage in Neuromorphic Nanoparticle Networks. *Nanotechnology* **2022**, *33* (27), 275602.

[49] Koo, S.; Park, J.; Koo, S.; Kim, K. Local Heat Dissipation of Ag Nanowire Networks Examined with Scanning Thermal Microscopy. *J. Phys. Chem. C* **2021**, *125* (11), 6306–6312.

[50] Kim, C. L.; Lee, J. Y.; Shin, D. G.; Yeo, J. S.; Kim, D. E. Mechanism of Heat-Induced Fusion of Silver Nanowires. *Sci. Rep.* **2020**, *10*, 9271.

[51] Sharvin, Y. V. A Possible Method for Studying Fermi Surfaces. *Sov. Phys. JETP* **1965**, *21*, 655.

[52] Wexler, G. The Size Effect and the Non-Local Boltzmann Transport Equation in Orifice and Disk Geometry. *Proc. Phys. Soc.* **1966**, *89* (4), 927–941.

[53] Torres, J. A.; Pascual, J. I.; Sáenz, J. J. Theory of Conduction through Narrow Constrictions in a Three-Dimensional Electron Gas. *Phys. Rev. B* **1994**, *49* (23), 16581–16584.

[54] López-Suárez, M.; Melis, C.; Colombo, L.; Tarantino, W. Modeling Charge Transport in Gold Nanogranular Films. *Phys. Rev. Mater.* **2021**, *5* (12), 126001.

[55] Erts, D.; Olin, H.; Ryen, L.; Thölén, A. Maxwell and Sharvin Conductance in Gold Point Contacts Investigated Using TEM-STM. *Phys. Rev. B - Condens. Matter Mater. Phys.* **2000**, *61* (19), 12725–12727.

[56] Ruzicka, J. Y.; Abu Bakar, F.; Hoeck, C.; Adnan, R.; Mcnicoll, C.; Kemmitt, T.; Cowie, B. C.; Metha, G. F.; Andersson, G. G.; Golovko, V. B. Toward Control of Gold Cluster Aggregation on TiO2 via Surface Treatments. *J. Phys. Chem. C* **2015**, *119* (43), 24465–24474.

[57] Verkhovtsev, A. V.; Erofeev, Y.; Solov'yov, A. V. Soft Landing of Metal Clusters on Graphite: A Molecular Dynamics Study. *Eur. Phys. J. D* **2020**, *74* (10), 205.

[58] Bardotti, L.; Jensen, P.; Hoareau, A.; Treilleux, M.; Cabaud, B.; Perez, A.; Aires, F. C. S. Diffusion and Aggregation of Large Antimony and Gold Clusters Deposited on Graphite. *Surf. Sci.* **1996**, *367* (3), 276–292.

[59] Solov'yov, I. A.; Yakubovich, A. V.; Nikolaev, P. V.; Volkovets, I.; Solov'yov, A. V. MesoBioNano Explorer-A Universal Program for Multiscale Computer Simulations of Complex Molecular Structure and Dynamics. *J. Comput. Chem.* **2012**, *33* (30), 2412–2439.

[60] Sushko, G. B.; Solov'yov, I. A.; Solov'yov, A. V. Modeling MesoBioNano Systems with MBN Studio Made Easy. *J. Mol. Graph. Model.* **2019**, *88*, 247–260.

[61] Gupta, R. P. Lattice Relaxation at a Metal Surface. *Phys. Rev. B* **1981**, *23* (12), 6265–6270.

[62] Cleri, F.; Rosato, V. Tight-Binding Potentials for Transition Metals and Alloys. *Phys. Rev. B* **1993**, *48* (1), 22–33.

[63] Brenner, D. W. Empirical Potential for Hydrocarbons for Use in Simulating the Chemical Vapor Deposition of Diamond Films. *Phys. Rev. B* **1990**, *42* (15), 9458–9471.

[64] Wang, W.; Wang, M.; Ambrosi, E.; Bricalli, A.; Laudato, M.; Sun, Z.; Chen, X.; Ielmini, D. Surface Diffusion-Limited Lifetime of Silver and Copper Nanofilaments in Resistive Switching Devices. *Nat. Commun.* **2019**, *10*, 81.

[65] Dick, V. V.; Solov'Yov, I. A.; Solov'Yov, A. V. Fragmentation Pathways of Nanofractal Structures on Surfaces. *Phys. Rev. B - Condens. Matter Mater. Phys.* **2011**, *84*, 115408.

[66] Woolard, E. W.; Einstein, A.; Furth, R.; Cowper, A. D. *Investigations on the Theory of the Brownian Movement*; 1928; Vol. 35.

[67] Antczak, G.; Ehrlich, G. *Surface Diffusion: Metals, Metal Atoms, and Clusters*; cambridge university press, 2010; Vol. 9780521899.

[68] Jensen, P.; Blase, X.; Ordejón, P. First Principles Study of Gold Adsorption and Diffusion on Graphite. *Surf. Sci.* **2004**, *564* (1–3), 173–178.

[69] Maruyama, Y. Temperature Dependence of Lévy-Type Stick-Slip Diffusion of a Gold Nanocluster on Graphite. *Phys. Rev. B - Condens. Matter Mater. Phys.* **2004**, *69* (24), 245408.

[70] El koraychy, E. yakout; Roncaglia, C.; Nelli, D.; Cerbelaud, M.; Ferrando, R. Growth Mechanisms from Tetrahedral Seeds to Multiply Twinned Au Nanoparticles Revealed by Atomistic Simulations. *Nanoscale Horizons* **2022**, *7* (8), 883–889.

[71] Nelli, D.; Rossi, G.; Wang, Z.; Palmer, R. E.; Ferrando, R. Structure and Orientation Effects in the Coalescence of Au Clusters. *Nanoscale* **2020**, *12* (14), 7688–7699.

[72] Pavloudis, T.; Kioseoglou, J.; Palmer, R. E. Bonding of Gold Nanoclusters on Graphene with and without Point Defects. *Nanomaterials* **2020**, *10* (11), 2109.

[73] Arcidiacono, S.; Bieri, N. .; Poulikakos, D.; Grigoropoulos, C. . On the Coalescence of Gold Nanoparticles. *Int. J. Multiph. Flow* **2004**, *30* (7–8), 979–994.

[74] Ibach, H.; Lüth, H. *Solid-State Physics*; Springer Berlin Heidelberg: Berlin, Heidelberg, 2009.

[75] Giovannetti, G.; Khomyakov, P. A.; Brocks, G.; Karpan, V. M.; Van Den Brink, J.; Kelly, P. J. Doping Graphene with Metal Contacts. *Phys. Rev. Lett.* **2008**, *101* (2), 026803.